\documentclass[preprint,aps]{revtex4}
\usepackage{array}
\usepackage{booktabs}
\usepackage{tabu}
\usepackage{dcolumn}
\usepackage{amsmath}
\usepackage{amsfonts}
\usepackage{amssymb}
\usepackage{graphicx}
\usepackage{subfigure}
\usepackage{graphicx}
\usepackage{dcolumn}
\usepackage{bm}
\usepackage{xcolor}

\def\be{\begin{equation}}
  \def\ee{\end{equation}}
\def\bea{\begin{eqnarray}}
\def\eea{\end{eqnarray}}
\def\f{\frac}
\def\n{\nonumber}
\def\l{\label}
\def\p{\phi}
\def\o{\over}
\def\R{\rho}
\def\pa{\partial}
\def\om{\omega}
\def\na{\nabla}
\def\P{\Phi}

\begin{document}

\title{Revisiting Scalar Tensor inflation by swampland criteria and Reheating}

\author{Abolhassan Mohammadi$^1$}
\email{a.mohammadi@uok.ac.ir; abolhassanm@gmail.com}
\author{Tayeb Golanbari$^1$}
\email{t.golanbari@uok.ac.ir; t.golanbari@gmail.com}
\author{Jamil Enayati$^2$}
\email{j.enayati@garmian.edu.krd}
\author{Shahram Jalalzadeh$^3$}
\email{Shahram@df.ufpe.br}
\author{Khaled Saaidi$^1$}
\email{ksaaidi@uok.ac.ir}

\affiliation{
$^1$Department of Physics, Faculty of Science, University of Kurdistan,  Sanandaj, Iran.\\
$^2$Physics Department, College of Education, University of Garmian, Iraq. \\
$^3$Departmento de Física, Universidade Federal de Pernambuco, Pernambuco, PE 52171-900, Brazil.
}
\date{\today}

\def\be{\begin{equation}}
  \def\ee{\end{equation}}
\def\bea{\begin{eqnarray}}
\def\eea{\end{eqnarray}}
\def\f{\frac}
\def\n{\nonumber}
\def\l{\label}
\def\p{\phi}
\def\o{\over}
\def\R{\rho}
\def\pa{\partial}
\def\om{\omega}
\def\na{\nabla}
\def\P{\Phi}

\begin{abstract}
The scenario of the slow-roll inflation is studied in the frame of the scalar-tensor theory of gravity where the scalar field has a non-minimal coupling to the geometric part. After deriving the main dynamical and perturbation equations, the model is investigated in detail for different functions of the potential and the coupling function. Considering the consistency of the model with observational model results into specific ranges for the free constants of the model. Then, the obtained ranges are reconsidered to realize whether they satisfy the swampland criteria. It is found that the swampland criteria impose another limitation and reduce the ranges. Using the final outcomes for the free constants of the model, we briefly consider the reheating phase to find out about the number of e-fold for the phase and the reheating temperature.
\end{abstract}
\maketitle

\section{Introduction}
The inflationary scenario is known as one of the best candidate for describing the very early universe evolution which has received a huge observational support \cite{Planck:2013jfk,Ade:2015lrj,Akrami:2019izv}. The scenario provides a clear explanation to the problems of the Big Bang theory by setting the initial conditions. The origin of the large scale of the universe is describe completely by the quantum fluctuation during inflation which is an amazing prediction of the inflationary scenario \cite{Linde:2005ht,Linde:2005vy,Linde:2004kg,Riotto:2002yw,Baumann:2009ds,Lyth:2009zz}. \\
The scenario has been extremely studied in the standard minimal coupling scalar field model which is the most simplest cosmological model. However, the inflation could be reestablished in the generalized models as
non-canonical \cite{Barenboim:2007ii,Franche:2010yj,Unnikrishnan:2012zu,Gwyn:2012ey,Rezazadeh:2014fwa,Cespedes:2015jga,Stein:2016jja,Saaidi:2015kaa}, tachyon inflation \cite{Fairbairn:2002yp,Mukohyama:2002cn,Feinstein:2002aj,Padmanabhan:2002cp,Aghamohammadi:2014aca},
DBI \cite{Spalinski:2007dv,Bessada:2009pe,Weller:2011ey,Nazavari:2016yaa,Amani:2018ueu},
G-inflation \cite{maeda2013stability,abolhasani2014primordial,alexander2015dynamics,tirandari2017anisotropic},
brane inflation \cite{maartens2000chaotic,golanbari2014brane,Mohammadi:2020ake},
and non-minimal coupling models of the scalar field. Also, it is possible to reconsider the scenario of inflation by assuming the existence of radiation, besides inflaton, which is known as warm inflation \cite{berera1995warm,berera2000warm,taylor2000perturbation,hall2004scalar,BasteroGil:2004tg,Sayar:2017pam,Akhtari:2017mxc,
Sheikhahmadi:2019gzs,Rasheed:2020syk}, or by relaxing the assumption of the smallness of the slow-roll parameters which is known as constant-roll inflation \cite{Motohashi:2014ppa,Gao:2018tdb,Odintsov:2017hbk,Oikonomou:2017bjx,Nojiri:2017qvx,
Motohashi:2019tyj,Odintsov:2019ahz,Mohammadi:2018oku,Mohammadi:2019dpu,Mohammadi:2018zkf,Mohammadi:2019qeu,
Mohammadi:2020ftb}. \\
After inflation, the universe is cold and empty of particles and all the energy is stored in homogeneous scalar field. The energy is required to transfer to the other matters to fulfill the universe, interact, and thermalize the universe in order to recover the big-bang nucleosynthesis. The scenario that explain this phenomenon is known as (p)reheating. The scenario explain the particle production after inflation, and clarify the final temperature of the universe after the phase and before the big-bang nucleosynthesis \cite{Kofman:1994rk,Kofman:1997yn,Bassett:2005xm,Allahverdi:2010xz,Amin:2014eta}. \\
Despite having an agreement with observational data, there are some theoretical tests that we desire to be satisfied by an inflationary model. The swampland criteria is one of these tests which recently has been introduced by \cite{Obied:2018sgi}, and then refined by \cite{Geng:2019phi,Ooguri:2018wrx}. If the swampland criteria is proved to be accurate, it could have a great impact of our understanding of the universe. The criteria are originated from string theory which is known as one of the best candidate for quantum gravity theory. There are many effective field theory (EFT) in which only some of them could formulate a consistent quantum gravity. These collection of EFTs are known as landscape. However, there are larger number of EFTs which do not lead to consistent quantum gravity, known as swampland. It includes two conjectures: 1) The first criterion is the distance conjecture which puts an upper bound on the range traversed by the scalar field in the field space, i.e. $\Delta \phi < c_1$ where $c_1$ is a constant of the order of one. 2) The second criterion is known as the de-Sitter conjecture which puts a lower bound on the gradient of the potential, i.e. $|V'/V| > c_2$ where $c_2$ is also a constant of the order of one \cite{Kehagias:2018uem}. It is the second criterion which has a direct tension with the slow-roll inflation scenario. The first slow-roll parameter is stated as $\epsilon = V'^2 / V^2$ which should be smaller than one during inflation. Then, it seems that the swampland criteria and the slow-roll inflation could not get alone. However, more investigation indicates that the generalized models of the inflation such as k-essence models, warm inflation, and brane inflation are able to survive and properly satisfy the criteria. \\
The main purpose of the present work is to consider the slow-roll inflation in the frame of the scalar-tensor theory of gravity where the scalar field has a non-minimal coupling to the geometric part of the theory. By comparing the model prediction with the observational data, a parametric space for the free constants of the model is achieved in which for every point in the space the model comes to a great consistency with data. The next goal is to consider the consistency of the model with the swampland criteria. It is believed that the inflation occurs in low-energy scale and it could be described by a low-energy EFT. Therefore, we are interested to build a model consistent with swampland criteria. More investigation indicates that our desire for satisfying the conjectures imposes more restriction on the free constants of the model and reduces the obtained parametric space. After inflation, the universe enters in reheating phase. Although the universe exits from the accelerated expansion phase, but it is still expanding. At the final step, the obtained results are utilize to realized that after the reheating time how much the universe expands, what is the final temperature of the universe, and for what equation of state we have a desirable outcomes.  \\
The paper is organized as follows: in Sec.II, the model is introduced and the main dynamical equations of the model are obtained. The cosmological perturbations is considered in Sec.III, where both scalar and tensor perturbations equations and the corresponding parameters are studied. The model was compared with observational data for three different choices of the potential and the coupling function, and the free constants of the model were determined. The limitation on the constants gets tighter after imposing the swampland criteria. A brief consideration about the reheating phase is presented in Sec.V, where we study the number of e-fold that the universe experience during the reheating phase and also the reheating temperature is obtained as well. The results are concluded in Sec.VI.

\section{The model}
The action for the model is defined as
\begin{equation}\label{action}
  S = \int d^4x \sqrt{-g} \left( {1 \over 2} \; f(\phi) \; R - \partial_\mu \phi \partial^\mu \phi - V(\phi) \right)
\end{equation}
where $g$ is the determinant of the metric $g_{\mu\nu}$ and $R$ is the Ricci scalar built based on this metric. The function $f(\phi)$ describes the coupling between the scalar field $\phi$ and the geometric part. Also, the potential of the scalar field is denoted by $V(\phi)$. \\
The field equations of the model are given by
\begin{eqnarray}
  3 f(\phi) H^2 &=& {1 \over 2} \; \dot{\phi}^2 + V(\phi) - 3H f'(\phi) \dot{\phi}, \label{Friedmann01} \\
  -2 f(\phi) \dot{H} &=& (1 + f''(\phi)) \; \dot{\phi}^2 + f'(\phi) (\ddot{\phi} - H \dot{\phi}), \label{Friedmann02} \\
  0 &=& \ddot{\phi} + 3 H \dot{\phi} + V'(\phi) - 3 f'(\phi) (2H^2 + \dot{H}). \label{EoM}
\end{eqnarray}
Taking $f(\phi)=1$, the equations come back to the canonical model of the scalar field. \\

The slow-roll parameters for the model are defined as following. There is the same definition for the first slow-roll parameter as the rate of the Hubble parameter during a Hubble time
\begin{equation}\label{SRP1}
  \epsilon_0 = - {\dot{H} \over H^2}
\end{equation}
which is assumed to be small during inflation. The second slow-roll parameter is defined based on the first one as
\begin{equation}\label{SRP2}
  \epsilon_1 = {\dot{\epsilon} \over H \epsilon}
\end{equation}
it should be noted that the second slow-roll parameter comes in other shape as well. Besides, in the scalar-tensor theory of gravity, there are other slow-roll parameters which two of them are expressed as
\begin{equation}\label{SRP34}
  \delta_0 = {\dot{f}(\phi) \over H f(\phi)}, \qquad \delta_1 = { \dot{\delta} \over H \delta}
\end{equation}

Substituting the field equations \eqref{Friedmann01}, \eqref{Friedmann02}, and \eqref{EoM}, one arrives at
\begin{equation}\label{dotphi}
  {1 \over 2} \; {\dot\phi^2 \over H^2 f(\phi)} = \epsilon_0 + {1 \over 2} \; \delta_0 - {\ddot{f}(\phi) \over 2 H^2 f(\phi)}
\end{equation}
where the last term is actually of the order of $\mathcal{O}(\epsilon^2)$ which could be proved by substituting the field equation in the slow-roll parameter $\delta_1$
\footnote{
the term is actually is obtained as ${\ddot{f}(\phi) / 2 H^2 f(\phi)} = \delta_0 (\delta_1 - \epsilon_0 + \delta_0)$
}.  \\
The potential of the model could be expressed in terms of the slow-roll parameters from Eq.\eqref{Friedmann01} as
\begin{equation}\label{potSRP}
  V(\phi) = H^2 f(\phi) \; \left( 3 - \epsilon_0 + {5 \over 2} \; \delta_0 + {1 \over 2} \; (\delta_1 - \epsilon_0 + \delta_0) \right).
\end{equation}

Assuming that the slow-roll parameters are smaller than one during inflation, the field equations could be simplified as follows
\begin{eqnarray}
  3 f(\phi) H^2 &=& V(\phi), \label{SimpFriedmann01}\\
  -2 f(\phi) \dot{H} &=& \dot{\phi}^2 - H \dot{f}(\phi), \label{SimpFriedmann02}\\
  0 &=& 3H \dot\phi + V'(\phi) - 6 f'(\phi) H^2. \label{SimpEoM}
\end{eqnarray}

\section{Cosmological Perturbations}
Cosmological perturbations are one of the most important prediction of the inflationary scenario. The perturbations are divided as scalar, vector, and tensor perturbations. Amongst them, the vector perturbation are ignored since they depend on the inverse of the scale factor and then are diluted during inflation. To study the cosmological perturbation of the model, it is required to calculate the action up to the second order of perturbations parameter. The most important perturbations are scalar and tensor perturbation which will be studied in the subsequent subsections.  \\

\subsection{Scalar perturbations}
The second-order action is given by \cite{Granda:2019wpe,DeFelice:2011zh,DeFelice:2011jm,Faraoni:2004pi}
\begin{equation}\label{2rdactionScalar}
  \delta S_s^{(2)} = \int d^4x a^3 \left[ \mathcal{G}_s \dot{\xi}^2 - {\mathcal{F}_s \over a^2} (\nabla \xi)^2 \right]
\end{equation}
where
\begin{equation*}
  \mathcal{G}_s = {\Sigma \over \Theta^2} f^2(\phi) + 3 f(\phi), \qquad
  \mathcal{F}_s = {1 \over a} \; {d \over dt}\left( {a \over \Theta} \; f^2(\phi) \right) - f(\phi)
\end{equation*}
and
\begin{equation*}
  \Theta = f(\phi) H - {1 \over 2} \dot{f}(\phi), \qquad
  \Sigma = -3 f(\phi) H^2 - 3 H \dot{f}(\phi) + {1 \over 2} \dot{\phi}^2
\end{equation*}
These parameters could also be rewritten in terms of the slow-roll parameters as
\begin{eqnarray}\label{ParamSRP}
  \Theta & = & f(\phi) H \left( 1 + {1 \over 2} \delta_0 \right), \\
  \Sigma & = & -f(\phi) H^2 \left( 3 - \epsilon_0 + {5 \over 2} \; \delta_0 + {\delta_0 \over 2} (\delta_1 - \epsilon_0 + \delta_0) \right), \nonumber \\
  \mathcal{G}_s & = & f(\phi) \; \left( 3 - {3 - \epsilon_0 + {5 \over 2} \; \delta_0 + {\delta_0 \over 2} (\delta_1 - \epsilon_0 + \delta_0) \over (1 + {1 \over 2} \delta_0)^2} \right), \nonumber \\
  \mathcal{F}_s & = & f(\phi) \left( {1 \over (1+{1 \over 2} \delta_0)} \; \left[ 1 + \delta_0 + \epsilon_0 - {1 \over 2} {\delta_0 \delta_1 \over (1+{1 \over 2} \delta_0)} \right] - 1 \right). \nonumber
\end{eqnarray}
Imposing the following variables change
\begin{equation*}
  d\tau_s = {c_s \over a} \; dt, \qquad
  z_s = \sqrt{2} \; a \left( \mathcal{F}_s \; \mathcal{G}_s \right)^{1/4}, \qquad
  v = z_s \xi
\end{equation*}
on the second-order action, it could be rearranged as
\begin{equation*}
  \delta S_s^{(2)} = {1 \over 2} \int d\tau_s d^3x \left( v'^2 - \nabla_i v \nabla_i v + {z''_s \over z_s} \; v \right).
\end{equation*}
Utilizing the Fourier representation for the parameter $v$,
\begin{equation*}
  v(\tau_s, x) = \int {d^3k \over (2\pi)^3} \; v_k(\tau_s) \; e^{ik.x}
\end{equation*}
and substituting it in the action, one has
\begin{equation*}
  \delta S_s^{(2)} = {1 \over 2} \int d\tau_s d^3x
\end{equation*}
which leads to the following differential equation for the perturbation parameter
\begin{equation}\label{vequation}
  v''_k + \left( k^2 - {z''_s \over z_s} \right) \; v_k = 0
\end{equation}
To find the solution, we first need to derive the term $z''_s / z_s$. From definition of the parameter $z_s$, and by using the slow-roll parameters it is obtained as
\begin{equation}\label{ZoverZ}
  {z''_s \over z_s} = {1 \over \tau_s^2} \; \left( \nu_s^2 - {1 \over 4} \right)
\end{equation}
in which the new parameter $\nu_s$ is given by
\begin{equation}\label{nu2}
  \nu_s^2 = {9 \over 4} \; \left( 1 + {4 \over 3} \; \epsilon_0 + {2 \over 3} \; \delta_0 +
  {2 \over 3} \; {2 \epsilon_0 \epsilon_1 + \delta_0 \delta_1 \over 2 \epsilon_0 + \delta_0} \right).
\end{equation}

The differential equation \eqref{vequation} in the subhorizon scales (i.e. $k \gg a H$), the equation reduces to a simple form, $v''_k + k^2 v_k =0$ with the following solution
\begin{equation}\label{subHsolution}
  v_k(\tau_s) = {1 \over \sqrt{2 k}} \; e^{-ik\tau_s}
\end{equation}
In general, the equation could be written as the Bessel differential equation by a variable change where the solution is first and second Hankel Function. Matching it with the subhorizon solution, only the first Hankel function is survived. Therefore, the general solution is given by
\begin{equation}\label{GeneralSolution}
  v_k(\tau_s) = {\sqrt{\pi} \over 2} \; e^{i{\pi \over 2}(\nu_s + 1/2)} \; \sqrt{-\tau_s} \; H_{\nu_s}^{(1)}(-k\tau_s)
\end{equation}
where up to the first order of the slow-roll parameter the parameter $\nu_s$ could be written as
\begin{equation}\label{nu}
  \nu_s = {3 \over 2} + \epsilon_0 + {1 \over 2} \delta_0 +
  {1 \over 2} {2 \epsilon_0 \epsilon_1 + \delta_0 \delta_1 \over 2 \epsilon_0 + \delta_0}
\end{equation}

The amplitude of the scalar perturbations is defined as
\begin{equation*}
  \mathcal{P}_s = {k^3 \over 2\pi^2} |\xi_k|^2
\end{equation*}
where $\xi = v_k / z_s$. To obtain the amplitude of the scalar perturbations on the superhorizon scales, first, it is required to evaluate $v_k(\tau_s)$ on scales $k \tau_s \gg 1$. After some manipulation, one arrives at
\begin{equation}\label{scalarPs}
  \mathcal{P}_s = {1 \over 8\pi^2} \; 2^{2\nu_s-3} \; \left( {\Gamma(\nu_s) \over \Gamma(3/2)} \right)^2 \;
                  {H^2 \over c_s^2 \sqrt{\mathcal{F}_s \mathcal{G}_s}} \; \left( -k \tau_s \right)^{3 - 2\nu_s}
\end{equation}
and the scalar spectral index is defined as
\begin{equation}\label{ns}
  n_s - 1 = {d\ln(\mathcal{P}_s \over d\ln(k)} = 3 - 2 \nu_s.
\end{equation}

\subsection{Tensor perturbations}
The second-order action for the tensor perturbations is given by \cite{Granda:2019wpe,DeFelice:2011zh,DeFelice:2011jm,Faraoni:2004pi}
\begin{equation}\label{2rdactionTensor}
  \delta S_t^{(2)} = \int d^4x a^2 \mathcal{G}_t \left[ \dot{h}^2_{ij} - {1 \over a^2} (\nabla h_{ij})^2 \right]
\end{equation}
Imposing the following variables changes
\begin{equation*}
  d\tau_t = {1 \over a} \; dt, \qquad
  z_t = {a \over 2} \; \sqrt{f(\phi)}, \qquad
  v_{ij} = z_t h_{ij}
\end{equation*}
on the action, and also applying the Fourier transformation leads to the following differential equation
\begin{equation}\label{vijequation}
  v''_{ij,k}(\tau_t) + \left( k^2 - {z''_t \over z} \right) \; v_{ij,k}(\tau_t) = 0
\end{equation}
which is the same equation as the scalar perturbations. To determine the solution, the term $z''_t/z_t$ should be specified which is given by
\begin{equation*}
  {z''_t \over z_t} = {\mu_t^2 - {1 \over 4} \over \tau_t^2}, \qquad
  \mu_t = {3 \over 2} + \epsilon_0 + {1 \over 2} \; \delta_0
\end{equation*}
The solution again is expressed in terms of the Hankel function, and regarding the fact that the amplitude of the tensor perturbation is defined as $\mathcal{P}_t = k^3 |h_{ij}|^2 / 2\pi^2$, one arrives at
\begin{equation}\label{pt}
  \mathcal{P}_t = {2 \over \pi^2} \; 2^{2\mu_t-3} \; \left( {\Gamma(\mu_t) \over \Gamma(3/2)} \right)^2 \;
                  {H^2 \over f(\phi)} \; \left( -k \tau_t \right)^{3 - 2\mu_t}
\end{equation}
and the tensor spectral index is read as
\begin{equation}\label{nt}
  n_t = 3 - 2\mu_t = -2\epsilon_0 - \delta_0
\end{equation}
The tensor perturbations are usually measured indirectly through the parameter $r$ which is defined as the ratio of the tensor perturbation to the scalar perturbations
\begin{equation}\label{r}
  r = {\mathcal{P}_t \over \mathcal{P}_s} = \epsilon_0 + {1 \over 2} \; \delta_0
\end{equation}

\section{Model consistency}
Here in this section we are going to study the model for different types of the potential and the coupling function. The cases will be considered in detail and their consistency with the observational data will be checked. By comparing the model with observational data, the valid parametric space for the free constants of the model is found out. Then, using these results for the constants, the validity of the swampland criteria is investigated, for all cases.

\subsection{First case}
\begin{figure}
  \centering
  \includegraphics[width=7cm]{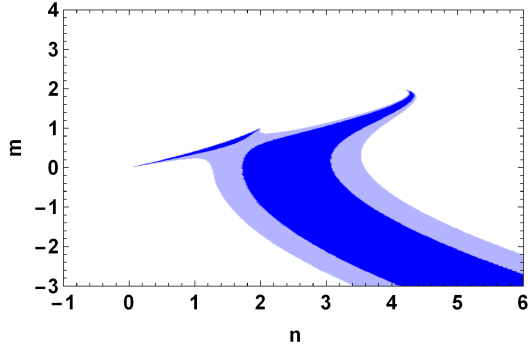}
  \caption{The parametric space of $n$ and $m$.}\label{FCnm}
\end{figure}

As the first case, both the potential and the coupling function is assumed as a power-law function of the scalar field
\begin{equation}\label{FC}
  V(\phi) = V_0 \phi^n, \qquad f(\phi) = f_0 \phi^m
\end{equation}
Substituting this function in Eqs.\eqref{SimpFriedmann01} and \eqref{SimpEoM}, the Hubble parameter and the time derivative of the scalar field are obtained
\begin{eqnarray}
  H^2 &=& {V_0 \over 3f_0} \; \phi^{n-m} \label{FChubble} \\
  \dot\phi &=& (2m-n) \; \sqrt{V_0 f_0 \over 3} \; \phi^{{n+m \over 2}-1} \label{FCphidot}
\end{eqnarray}
Combining these two equations, we have
\begin{equation}\label{FCHdot}
  \dot{H} = {(n-m)\; (2m-n) \over 6} \; V_0 \phi^{n-2}
\end{equation}
The slow-roll parameters are also obtained in terms of the scalar field as follows
\begin{eqnarray}
  \epsilon_0(\phi) &=& {(n-m) \; (n-2m) \over 2} \; f_0 \phi^{m-2}, \label{FCSRP} \\
  \epsilon_1(\phi) &=& (m-2) \; (2m-n) \; f_0 \phi^{m-2}, \nonumber \\
  \delta_0(\phi) &=& m(2m-n) \; f_0 \phi^{m-2} \nonumber \\
  \delta_1(\phi) &=& m(m-2)(2m-n) f_0 \phi^(m-2). \nonumber
\end{eqnarray}
The inflation ends as the first slow-roll parameter $\epsilon_0$ reaches unity where the scalar field is read as
\begin{equation}\label{FCphiend}
  \phi_e^{m-2} = {2 \over (n-m)(n-2m) \; f_0}.
\end{equation}
We are more interested in the values of the parameters at the time of the horizon crossing. In this regard, we first need to obtain the scalar field at the horizon crossing time which is computed through the number of e-fold, as
\begin{equation}\label{FCphistar}
  \phi_\star^{m-2} = {2 \over (n-m)(n-2m) \; f_0} \;
                     \left( 1 + {2(2-m) \over n-m} \; N \right)^{-1}
\end{equation}
Using this result, the slow-roll parameters at the horizon crossing time are
\begin{eqnarray}
  \epsilon_0^\star &=& \left( 1 + {2(2-m) \over n-m} \; N \right)^{-1} \label{FCSRPstar} \\
  \epsilon_1^\star &=& {-2 (m-2) \over (n-m)} \; \epsilon_0^\star \nonumber \\
  \delta_0^\star &=& {-2m \over (n-m)} \; \epsilon_0^\star \nonumber \\
  \delta_1^\star &=& {-2m(m-2) \over (n-m)} \; \epsilon_0^\star. \nonumber
\end{eqnarray}
The slow-roll parameters at the horizon crossing time depend on the free parameters of the model, i.e. $n$ and $m$. Then, from Eqs.\eqref{ns} and \eqref{r}, it is realized that the scalar spectral index and the ratio of tensor-to-scalar perturbations also depend on these two parameters as
\begin{eqnarray}\label{FCrns}
  n_s^\star - 1 & = & \left( -2 + {2m \over n-m} + {4(m-2)(n-2m) \over 2(n-m)(n-2m)} \right) \; \epsilon_0^\star\\
  r^\star & = & \left( 1 - {m \over n-m} \right) \; \epsilon_0^\star \nonumber
\end{eqnarray}
Utilizing the Planck $r-n_s$ diagram, we could determine the values of $(n,m)$ in which to get values for $n_s$ and $r$ consistent with observation. The set of these $(n,m)$ points creates a parametric space which is plotted in Fig.\ref{FCnm}. \\
The other two free parameters $V_0$ and $f_0$ are determined using the observational data for the amplitude of the scalar perturbations and energy scale of inflation which specifies by the value of the potential at the horizon crossing time. Computing the potential at the horizon crossing using Eq.\eqref{FCphistar}, the parameter $V_0$ is read as
\begin{equation}\label{FCV0}
  V_0 = \left( 2 \epsilon_0^\star f_0 \over (n-m)(n-2m) \right)^{n \over m-2} \; V^\star
\end{equation}
where $V^\star$ is the energy scale of inflation. Applying Eqs.\eqref{FChubble}, \eqref{FCphistar}, and \eqref{FCV0} on amplitude of the scalar field \eqref{scalarPs}, the parameter $f_0$ is found to be
\begin{equation}\label{FCf0}
  f_0^{m+2 \over m-2} = {\mathcal{P}_s^\star \over \left( 2 \epsilon_0^\star \over (n-m)(n-2m) \right)^{n \over m-2} \; {(n-m)V^\star \over 8\pi^2 (n-2m)\epsilon_0^\star}
  \left[ {2 \epsilon_0^\star \over (n-m)(n-2m)} \right]^{n-2m \over m-2} }
\end{equation}
where $\mathcal{P}_s^\star$ is the values of the amplitude of the scalar perturbations at the horizon crossing taken from observational data of Planck.

\begin{figure}
  \centering
  \includegraphics[width=7cm]{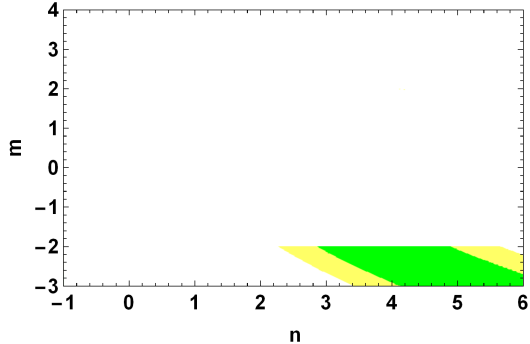}
  \caption{The parametric space of $n$ and $m$.}\label{FCnmSC}
\end{figure}

So far we consider the consistency of the model with observational data, and we could determine a range for the free parameters of the model which produce the agreement. However, there are swampland criteria which could impose anther constraint on the model. The swampland criteria includes two parts: first, there is an upper bound on the scalar field range in which $\Delta\phi < c_1$, and the second criterion puts a lower band on the gradient of the potential so that $|V'/V| > c_2$. Our desire for satisfying these two conjecture put another restriction of the free parameters of the model. From Eqs.\eqref{FC}, \eqref{FCphiend}, and \eqref{FCphistar} it is easily concluded that these two conjecture could be reexpressed in terms of the free parameters of the model. Regarding this, we could add two other conditions to find the actual parameter space which simultaneously brings a consistency with observational data and also satisfy the swampland criteria. Fig.\ref{FCnmSC} displays this parametric space for the parameter $n$ and $m$.

\begin{figure}
  \centering
  \subfigure[\label{FCdphi}]{\includegraphics[width=7cm]{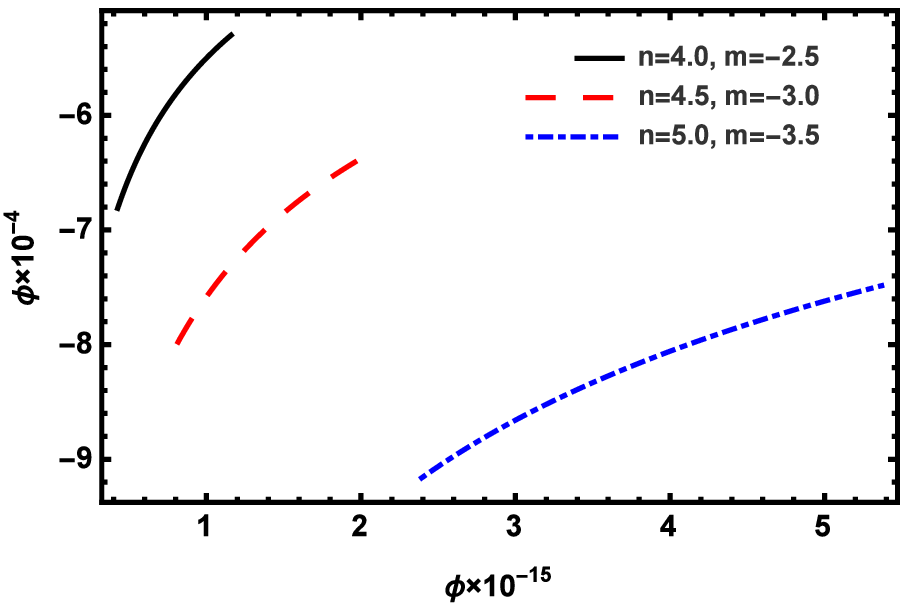}}
  \subfigure[\label{FCepsilon}]{\includegraphics[width=7cm]{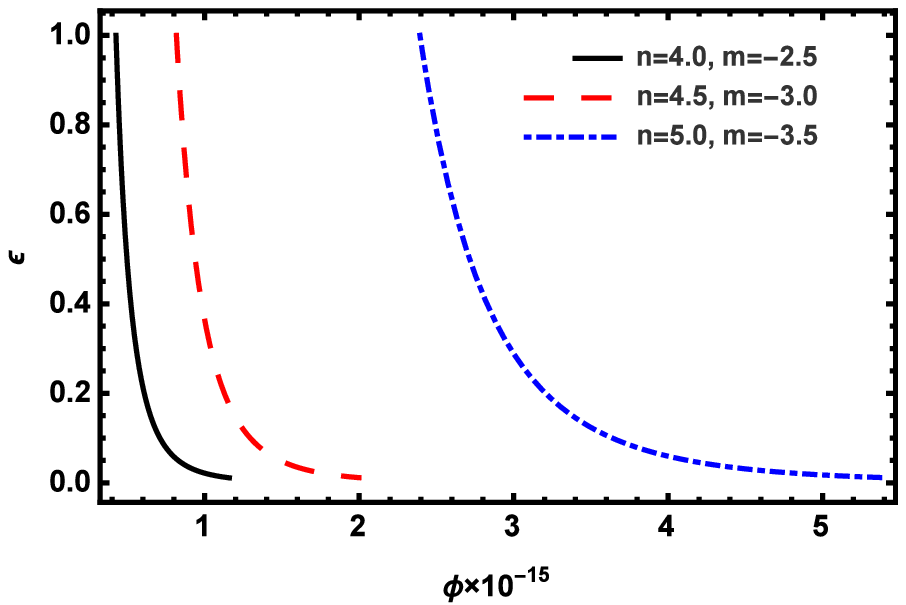}}
  \caption{Behavior of the a) time derivative of the scalar field, b) first slow-roll parameter are plotted versus the scalar field for three different values of $n$ and $m$.}\label{FCdphiepsilon}
\end{figure}

\begin{figure}
  \centering
  \includegraphics[width=7cm]{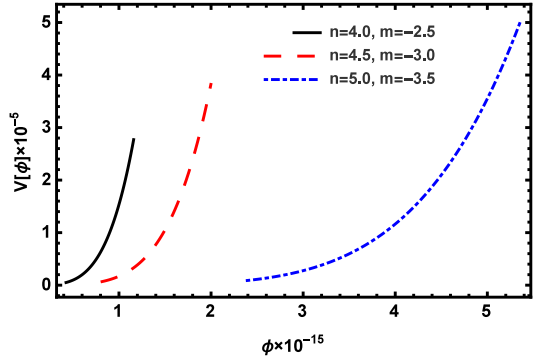}
  \caption{The potential versus the scalar field in inflationary times for three different values of $n$ and $m$.}\label{FCpotential}
\end{figure}

The numerical results, presented in Table.\ref{FCtable}, gives some insight about the result of the case. For each pair of $(n,m)$, selected from parametric area of Fig.\ref{FCnmSC}, one could determine all model parameters and also the perturbation parameters. The result indicates that the obtained $n_s$ and $r$ are in good agreement with data, and the obtained values for the terms $\Delta\phi$ and $|V'/V|$ perfectly satisfy the conditions of the swampland. In the last two rows of the table, a pair of $(n,m)$ is selected in which it stands in the parametric area of Fig.\ref{FCnm} but not in the parametric area of Fig.\ref{FCnmSC}. It is found out that the result of the model agrees with observational data, however it violates both swampland criteria. \\
\begin{table}
  \centering
  \begin{tabular}{p{1cm}p{1.2cm}p{2.5cm}p{2.5cm}p{1.5cm}p{1.5cm}p{2.5cm}p{2.5cm}}
    \hline
    \ $n$ & \ \ $m$ & \qquad $f_0$ & \qquad $V_0$ & \ \ $n_s$ & \quad $r$ & \ \quad $\Delta\Phi$ & \quad $|V'/V|$ \\

    \hline

    $3.7$ & $-3.0$ & $9.22 \times 10^{-40}$ & $1.92 \times 10^{21}$ & $0.9741$ & $0.0147$ & $4.69 \times 10^{-8}$ & $4.72 \times 10^{7}$ \\

    $4.5$ & $-3.0$ & $4.22 \times 10^{-43}$ & $1.69 \times 10^{30}$ & $0.9702$ & $0.0159$ & $1.18 \times 10^{-8}$ & $2.25 \times 10^{9}$ \\

    $5.5$ & $-3.0$ & $1.44 \times 10^{-46}$ & $6.65 \times 10^{42}$ & $0.9657$ & $0.0174$ & $2.05 \times 10^{-9}$ & $1.56  \times 10^{7}$ \\

    $4.0$ & $-2.5$ & $2.31 \times 10^{-71}$ & $1.53 \times 10^{55}$ & $0.9691$ & $0.0152$ & $7.34 \times 10^{-16}$ & $3.45 \times 10^{15}$ \\

    $5.0$ & $-2.5$ & $9.75 \times 10^{-79}$ & $1.19 \times 10^{79}$ & $0.9643$ & $0.0168$ & $1.69 \times 10^{-17}$ & $1.83  \times 10^{17}$ \\

    \hline

    $3.5$ & $-1.5$ & $3.03 \times 10^{-48}$ & $8.24 \times 10^{-57}$ & $0.9653$ & $0.0141$ & $4.16 \times 10^{14}$ & $6.09 \times 10^{-15}$ \\

    $2.5$ & $-1.0$ & $4.25 \times 10^{16}$ & $3.47 \times 10^{-22}$ & $0.9686$ & $0.0114$ & $2.65 \times 10^{6}$ & $7.46 \times 10^{-7}$ \\
    \hline
  \end{tabular}
  \caption{table}\label{FCtable}
\end{table}
After determining the free constants of the model, we can now consider the consistency of the model. Fig.\ref{FCdphiepsilon} displays the behavior of the time derivative of the scalar field and the first slow-roll parameter for three different choices of the constants $n$ and $m$. From Fig.\ref{FCdphi} it is clear that for the case the parameter $\dot{\phi}$ is always negative during inflation which indicates that the scalar field decreases by passing time and reaching end of inflation. Therefore, it is expected that the slow-roll parameter $\epsilon_1$ increases by reducing the scalar field.  Behavior of $\epsilon$ versus the scalar field is portrayed in Fig.\ref{FCepsilon} during inflationary time. The parameter increases by reducing the scalar field and it reaches one indicating the end of inflation. \\
The behavior of the potential is depicted in Fig.\ref{FCpotential} versus the scalar field for different values of $n$ and $m$ where larger values of the scalar field indicates earlier times of inflation. The potential has larger values for bigger values of the scalar field and it reduces by decreasing of the scalar field. This is the usual behavior for the potential in which the scalar field rolls down from the top to the minimum of the potential.

\subsection{Second case}
\begin{figure}
  \centering
  \includegraphics[width=7cm]{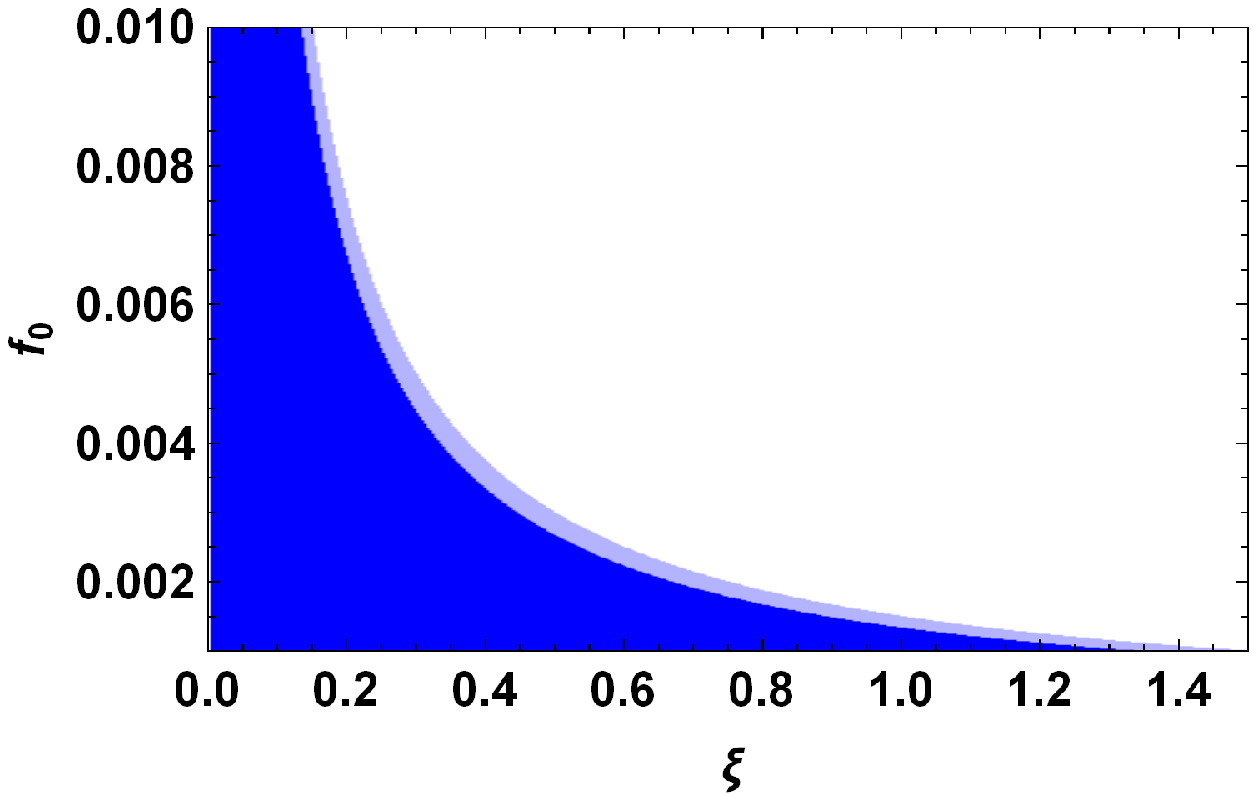}
  \caption{The parametric space of $\xi$ and $f_0$.}\label{SCxf}
\end{figure}

\begin{figure}
  \centering
  \includegraphics[width=7cm]{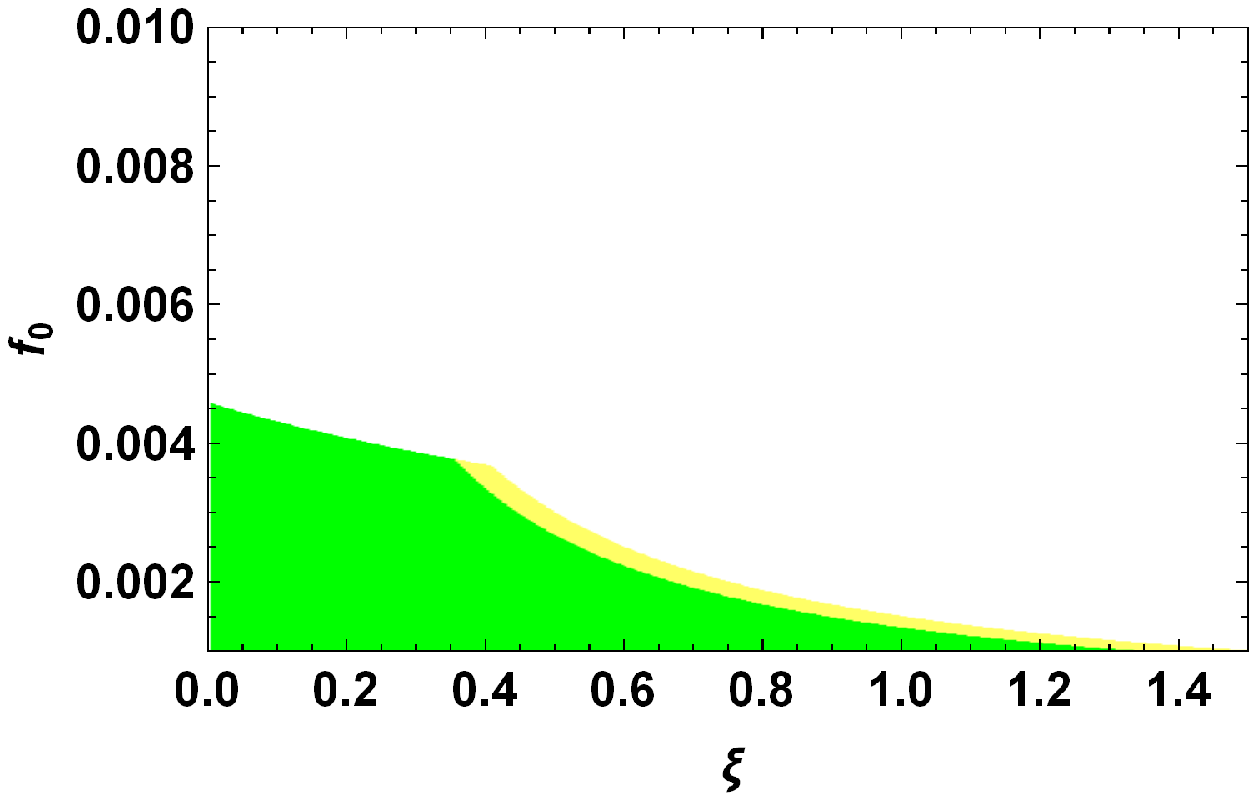}
  \caption{The parametric space of $\xi$ and $f_0$.}\label{SCxfSC}
\end{figure}

For this case, the potential of the scalar field and the coupling function are taken as follows
\begin{equation}\label{SC}
  V(\phi) = {1 \over 2} \; m^2 \phi^2, \qquad f(\phi) = f_0 (1 - \xi \phi^2)
\end{equation}
Applying these definitions, the Hubble parameter and the time derivative of the scalar field are read from Eqs.\eqref{SimpFriedmann01} and \eqref{SimpEoM} as
\begin{eqnarray}
  H^2 &=& {m^2 \over 6 f_0} \; {\phi^2 \over 1-\xi \phi^2}, \label{SChubble}\\
  \dot\phi &=& - \sqrt{2m^2 f_0 \over 3} \; {1+\xi \phi^2 \over \sqrt{1-\xi\phi^2}}. \label{SCphidot}
\end{eqnarray}
Using above equation, the time derivative of the Hubble parameter is obtained as a function of the scalar field
\begin{equation}\label{SCHdot}
  \dot{H} = {-m^2 \over 3} \; {1+\xi\phi^2 \over (1-\xi\phi^2)^2}
\end{equation}
With these results, the slow-roll parameters are simply extracted in which
\begin{eqnarray}
  \epsilon_0(\phi) &=& 2f_0 \; {(1+\xi\phi^2) \over \phi^2 (1-\xi\phi^2)}, \label{SCSRP} \\
  \epsilon_1(\phi) &=& 4f_0 \; {1 - \xi\phi^2 (2+\xi\phi^2) \over \phi^2 (1-\xi\phi^2)}, \nonumber \\
  \delta_0(\phi) &=& 4f_0 \; {\xi (1+\xi\phi^2) \over (1-\xi\phi^2)}, \nonumber \\
  \delta_1(\phi) &=& -8f_0 \; {\xi \over (1-\xi\phi^2)}. \nonumber
\end{eqnarray}
Inflation ends as the first slow-roll parameter reaches unity, where the scalar field is read
\begin{equation}\label{SCphiend}
  \xi\phi^2 = {1 \over 2} \; \left[ (1-2f_0\xi) \pm \sqrt{1-12f_0\xi + 4f_0^2 \xi^2} \right]
\end{equation}
Substituting this result in the number of e-fold, one could read the scalar field at the earlier times where the perturbations cross the horizon. The scalar field at this time is acquired in terms of the number of e-fold
\begin{equation}\label{SCphistar}
  (1+\xi\phi^2_star) = (1+\xi\phi^2_e) \; e^{4f_0 \xi N}.
\end{equation}
Then, the slow-roll parameter at the time of the horizon crossing also obtained in terms of the number of e-fold and the free constants of the model
\begin{eqnarray}
  \epsilon_0^\star &=& 2f_0 \xi {(1+\xi\phi_e^2) \; e^{4f_0 \xi N} \over \left[ (1+\xi\phi_e^2) \; e^{4f_0 \xi N} - 1 \right] \; \left[ 2 - (1+\xi\phi_e^2) \; e^{4f_0 \xi N} \right]} \label{SCSRPstar} \\
  \epsilon_1^\star &=& 4f_0 \xi \; {1 - \left( (1+\xi\phi_e^2) \; e^{4f_0 \xi N} - 1 \right) \; \left( 1 + (1+\xi\phi_e^2) \; e^{4f_0 \xi N} \right) \over \left[ (1+\xi\phi_e^2) \; e^{4f_0 \xi N} - 1 \right] \; \left[ 2 - (1+\xi\phi_e^2) \; e^{4f_0 \xi N} \right]} \nonumber \\
  \delta_0^\star &=& 4f_0 \xi \; {(1+\xi\phi_e^2) \; e^{4f_0 \xi N} \over 2 - (1+\xi\phi_e^2) \; e^{4f_0 \xi N}}, \nonumber \\
  \delta_1^\star &=& -8f_0 \xi \; {1 \over 2 - (1+\xi\phi_e^2) \; e^{4f_0 \xi N}}. \nonumber
\end{eqnarray}
Then, from Eqs.\eqref{ns} and \eqref{r} it could be concluded that the scalar spectral index and the tensor-to-scalar ratio at the time of the horizon crossing depend only on the free constants of the model. Therefore, by applying the $r-n_s$ diagram of the Planck one could determine the acceptable range for the constants $\xi$ and $f_0$ in which for any $(\xi, f_0)$ point in the range, the model comes to a proper agreement with observational data. The parameter space is plotted in Fig.\ref{SCxf} which shows the acceptable range for the constants $\xi$ and $f_0$.  \\
The constant $m$ is the last parameter to be determined. Using the observational data for the amplitude of the scalar perturbations, the constant $m$ could be read as
\begin{equation}\label{m}
  m^2 = \left( 48 \pi^2 f_0^2 \xi \mathcal{P}_s^\star \right) \;
  {(2 - (1+\xi\phi_e^2) \; e^{4f_0 \xi N}) \; (1+\xi\phi_e^2) \; e^{4f_0 \xi N} \; \epsilon_0^\star
  \over (1+\xi\phi_e^2) \; e^{4f_0 \xi N} -1}
\end{equation}
This parameter with the other constants $\xi$ and $f_0$ are utilized to specify the energy scale of inflation
\begin{equation}\label{SCV0}
  V^\star = {m^2 \over 2} \; {(1+\xi\phi_e^2) \; e^{4f_0 \xi N}-1 \over \xi}.
\end{equation}
Fig.\ref{SCxf} displays the parametric space with bring an agreement with observational data. However, there will be more restriction by adding the swampland criteria on the model. Eqs.\ref{SCphiend} and \eqref{SCphistar} are used to build the first swampland criteria which is the field distance, and the potential \eqref{SC} and \eqref{SCphistar} are utilized to rewrite the relation correspond to the second swampland criterion. Satisfying two conditions impose more restriction on the parametric space in which Fig.\ref{SCxfSC} the refined parametric space. For any $(\xi,f_0)$ in this space the model has an agreement with observational data and also perfectly satisfies the swampland criteria.

\begin{figure}
  \centering
  \subfigure[\label{SCdphi}]{\includegraphics[width=7cm]{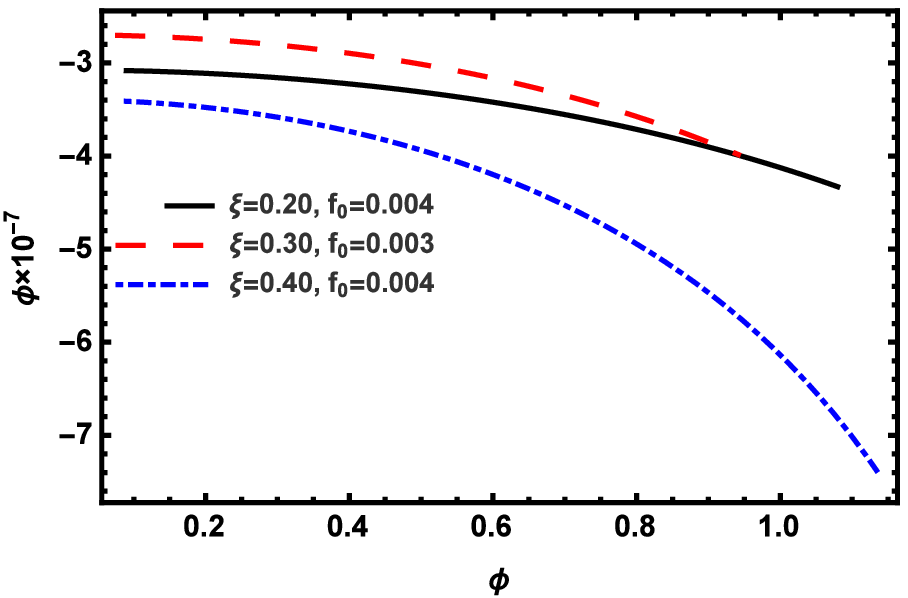}}
  \subfigure[\label{SCepsilon}]{\includegraphics[width=7cm]{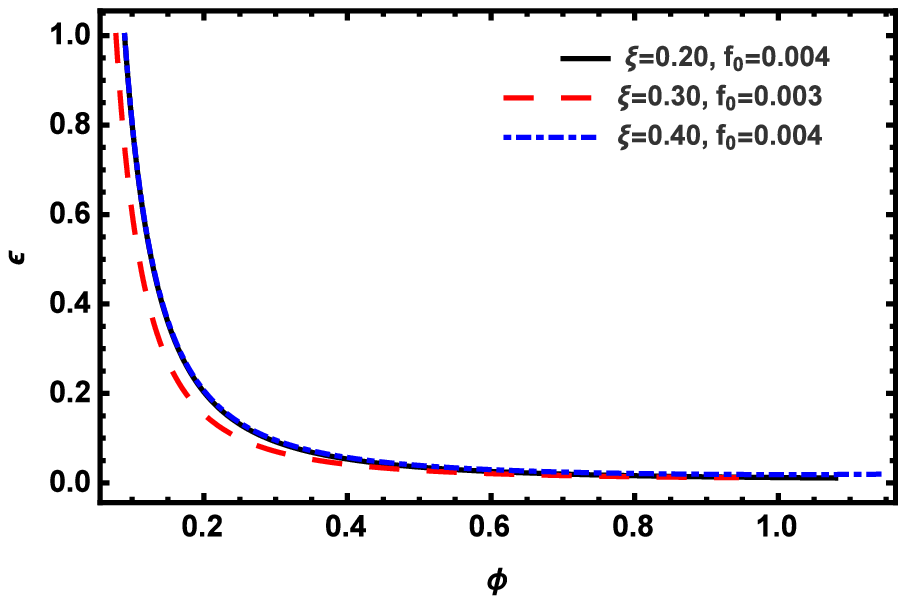}}
  \caption{Behavior of the a) time derivative of the scalar field, b) first slow-roll parameter are plotted versus the scalar field for three different values of $\xi$ and $f_0$.}\label{SCdphiepsilon}
\end{figure}

\begin{figure}
  \centering
  \includegraphics[width=7cm]{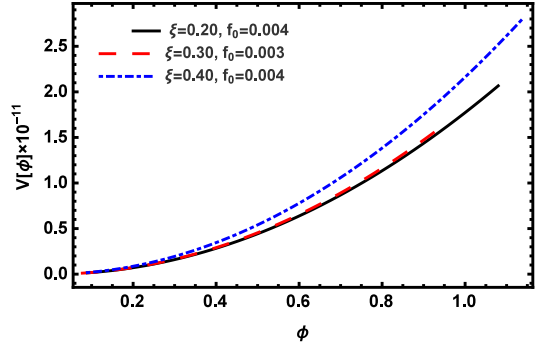}
  \caption{The potential versus the scalar field in inflationary times for three different values of $\xi$ and $f_0$.}\label{SCpotential}
\end{figure}

The numerical results for the case are prepared in Table.\ref{SCtable}. For each pair of $(\xi,f_0)$ taken from Fig.\ref{SCxfSC}, the other parameters are derived where the constant $m$ is of the order of $10^{-6}$ and the energy scale of inflation found to be of the order of $10^{-3}(M_p)$. The results for the $n_s$ and $r$ demonstrate that the model is in good agreement with observational data. Also, the last two columns state that the two swampland criteria are satisfied for the selected values of $\xi$ and $f_0$. For the last two rows, the $(\xi,f_0)$ points are taken from the parametric space of Fig.\ref{SCxf} and they do not stand in the parametric space of Fig.\ref{SCxfSC}. For these choices, the model satisfies the second swampland criterion, however it violates the first swampland conjecture regarding the field distance. \\
\begin{table}
  \centering
  \begin{tabular}{p{1.2cm}p{1.2cm}p{2.5cm}p{2.5cm}p{1.5cm}p{1.5cm}p{1.3cm}p{1.3cm}}
    \hline
    \ \ $\xi$ & \ \ $f_0$ & \qquad $m$ & \qquad $V^\star$ & \quad $n_s$ & \quad $r$ & $\Delta\phi$ & $|V'/V|$ \\

    \hline

    $0.20$ & $0.002$ & $5.65 \times 10^{-6}$ & $8.84 \times 10^{-12}$ & $0.9690$ & $0.0100$ & $0.67$ & \ $2.69$ \\

    $0.20$ & $0.004$ & $5.95 \times 10^{-6}$ & $2.06 \times 10^{-11}$ & $0.9674$ & $0.0136$ & $0.99$ & \ $1.85$ \\

    $0.25$ & $0.003$ & $5.91 \times 10^{-6}$ & $1.52 \times 10^{-11}$ & $0.9677$ & $0.0130$ & $0.85$ & \ $2.14$ \\

    $0.40$ & $0.002$ & $5.93 \times 10^{-6}$ & $1.03 \times 10^{-11}$ & $0.9672$ & $0.0136$ & $0.70$ & \ $2.62$ \\

    $0.70$ & $0.001$ & $5.87 \times 10^{-6}$ & $4.47 \times 10^{-11}$ & $0.9679$ & $0.0125$ & $0.49$ & \ $3.73$ \\

    \hline

    $0.30$ & $0.005$ & $6.49 \times 10^{-6}$ & $3.38 \times 10^{-11}$ & $0.9579$ & $0.0263$ & $1.16$ & \ $1.57$ \\

    $0.20$ & $0.006$ & $6.26 \times 10^{-6}$ & $3.62 \times 10^{-11}$ & $0.9636$ & $0.0193$ & $1.25$ & \ $1.47$ \\

    \hline
  \end{tabular}
  \caption{table}\label{SCtable}
\end{table}
The best choices of the free constants of $\xi$ and $f_0$ is displayed in Fig.\ref{SCxfSC}. Using these values, one could plot the behavior of the parameter $\dot\phi$ and $\epsilon_1$ versus the scalar field, i.e. Egs.\eqref{SCphidot} and \eqref{SCSRP}, during inflation. The time derivative of the scalar field is negative during the inflationary times, presented in Fig.\ref{SCdphi}, which indicates that the scalar field has bigger values at the beginning of inflation and it reduces by reaching the end of inflation. Then, the first slow-roll parameter $\epsilon$ gets smaller values for the bigger values of the scalar field and it gets larger for the smaller scalar field and finally reaches one and inflation ends. This behavior is clearly presented in Fig.\ref{SCepsilon}. \\
The potential of the case is presented in Fig.\ref{SCpotential} for different choices of $\xi$ and $f_0$. The potential stands in larger values at the beginning of inflation and it reduces by passing time and approaching to the end of inflation.

\subsection{Third case}
\begin{figure}
  \centering
  \includegraphics[width=7cm]{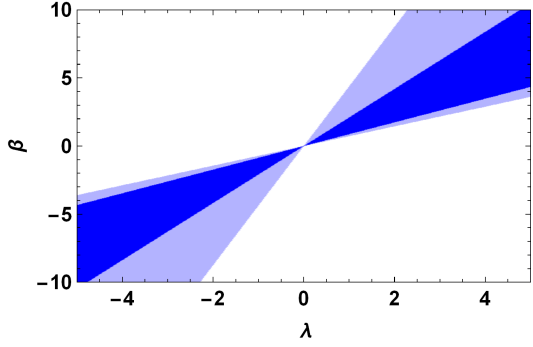}
  \caption{The parametric space of $\beta$ and $\lambda$.}\label{TClb}
\end{figure}

\begin{figure}
  \centering
  \includegraphics[width=7cm]{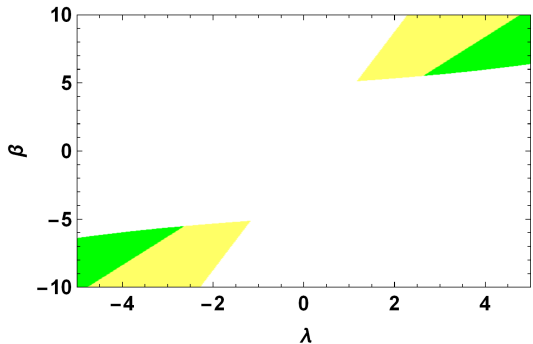}
  \caption{The parametric space of $\beta$ and $\lambda$.}\label{TClbSC}
\end{figure}

For the last case, we consider an exponential function of the scalar field for both the scalar field potential and the coupling function,
\begin{equation}\label{TC}
  V(\phi) = V_0 \; e^{\lambda \phi}, \qquad f(\phi) = f_0 \; e^{\beta \phi}.
\end{equation}
Following Eqs.\eqref{SimpFriedmann01} and \eqref{SimpEoM}, the Hubble parameter and the time derivative of the scalar field are obtained in terms of the scalar field as well
\begin{eqnarray}
  H^2 &=& {V_0 \over 3 f_0} \; e^{(\lambda-\beta)\phi}, \label{TChubble} \\
  \dot{\phi} &=& - \sqrt{V_0 f_0 \over 3} \; (\lambda-2\beta) \; e^{{(\lambda+\beta) \over 2} \; \phi}, \label{TCphidot}
\end{eqnarray}
Taking time derivative of the Hubble parameter and using Eq.\eqref{TCphidot}, the time derivative of the Hubble parameter is derived also as a function of the scalar field
\begin{equation}\label{TCHdot}
  \dot{H} = - {(\lambda - \beta)(\lambda-2\beta) \over 6} \; V_0 \; e^{\lambda \phi}.
\end{equation}
Substituting these results in Eq.\eqref{SRP1}, \eqref{SRP2}, and \eqref{SRP34}, the slow-roll parameters are given by
\begin{eqnarray}
  \epsilon_0(\phi) &=& {1 \over 2} \; f_0 \; (\lambda-\beta) (\lambda-2\beta) \; e^{\beta \phi}, \label{SCSRP} \\
  \epsilon_1(\phi) &=& -\beta f_0 \; (\lambda-2\beta) \; , \nonumber \\
  \delta_0(\phi) &=& \delta_1(\phi) = \epsilon_1(\phi), \nonumber \\
\end{eqnarray}
As the first slow-roll parameter $\epsilon_0$ reaches one the inflation ends. The scalar field at this time is
\begin{equation}\label{TCphiend}
  e^{\beta \phi_e} = {2 \over f_0 \; (\lambda-\beta) (\lambda-2\beta)}.
\end{equation}
However, we are more interested to find the parameters at the time when the perturbations cross the horizon. in this regard, we first obtain the scalar field at the time of the horizon crossing which is performed through the number of e-fold equation and by using Eq.\eqref{TCphiend}, so that
\begin{equation}\label{TCphistar}
  e^{\beta \phi_\star} = e^{\beta \phi_e} \; \left( 1 - {2\beta \over (\lambda-\beta)} \; N \right)^{-1} .
\end{equation}
Since all the main dynamical and perturbation parameters are expressed in terms of the scalar field, one could easily evaluate them at the time of the horizon crossing. As the first step, we express the slow-roll parameters at the time
\begin{eqnarray}
  \epsilon_0^\star &=& \left( 1 - {2\beta \over (\lambda-\beta)} \; N \right)^{-1} \label{SCSRPstar} \\
  \epsilon_1^\star &=& \delta_0^\star = \delta_1^\star = {-2\beta \over (\lambda - \beta)} \; \epsilon_0^\star \nonumber \\
\end{eqnarray}
Substituting the above slow-roll parameters in Eqs.\eqref{ns} and \eqref{r}, the scalar spectral index and the tensor-to-scalar ratio are estimated at the time of the horizon crossing
\begin{eqnarray}
  n_s^\star &=& 1 - {2 (\lambda-3\beta) \over (\lambda-\beta)} \; \epsilon_0^\star \\
  r^\star &=& {(\lambda-2\beta) \over (\lambda-\beta)}
\end{eqnarray}
which leads to this conclusion that these two parameters only depend on the model constants $\beta$ and $\lambda$. By comparing the results of the model about $n_s^\star$ and $r^\star$ for different values of $\lambda$ and $\beta$, and comparing them with $r-n_s$ diagram of Planck, leads us to an acceptable range for the both parameter in which for every $(\lambda,\beta)$ point in the range the result of the model comes to a good agreement with observational data. This acceptable range is depicted in Fig.\ref{TClb}. \\
The other two constants, i.e. $V_0$ and $f_0$, are determined using the observational data for the amplitude of the scalar perturbations and energy scale of inflation. From the relation of the amplitude of the scalar perturbation and estimating the parameter at the time of the horizon crossing, one arrives at
\begin{equation}\label{TCf0}
  {f_0^{\lambda/\beta} \over V_0} = {(\lambda-\beta)  \over 24\pi^2 \mathcal{P}_s^\star (\lambda-2\beta) \epsilon_0^\star} \;
                      \left( 2 \epsilon_0^\star \over (\lambda-\beta)(\lambda-2\beta) \right)^{{\lambda \over \beta}-2}
\end{equation}
The term on the left hand side of the above equation exactly appears in the potential, and then one could determine the energy scale of the inflation as
\begin{equation}\label{TCV)}
  V^\star = {V_0 \over f_0^{\lambda/\beta}} \;
                \left( 2 \epsilon_0^\star \over (\lambda-\beta)(\lambda-2\beta) \right)^{\lambda \over \beta}
\end{equation}
Using the observational data, we could determine a parametric space for the free constants of the model. However, they could be more restricted using the swampland criteria. Using the Eqs.\eqref{TC},\eqref{TCphiend}, and \eqref{TCphistar}, the swampland criteria, i.e. $\Delta\phi < 1 $ and $|V'/V| > 1$, are rewritten in terms of the constants of the model. Therefore, the desire for satisfying these two conjectures imposes another restriction besides our main attention for having an agreement with data. The refined parametric space is shown in Fig.\ref{TClbSC} which clearly displays that the space has been restricted with respect to Fig.\ref{TClb}.

\begin{figure}
  \centering
  \subfigure[\label{TCdphi}]{\includegraphics[width=7cm]{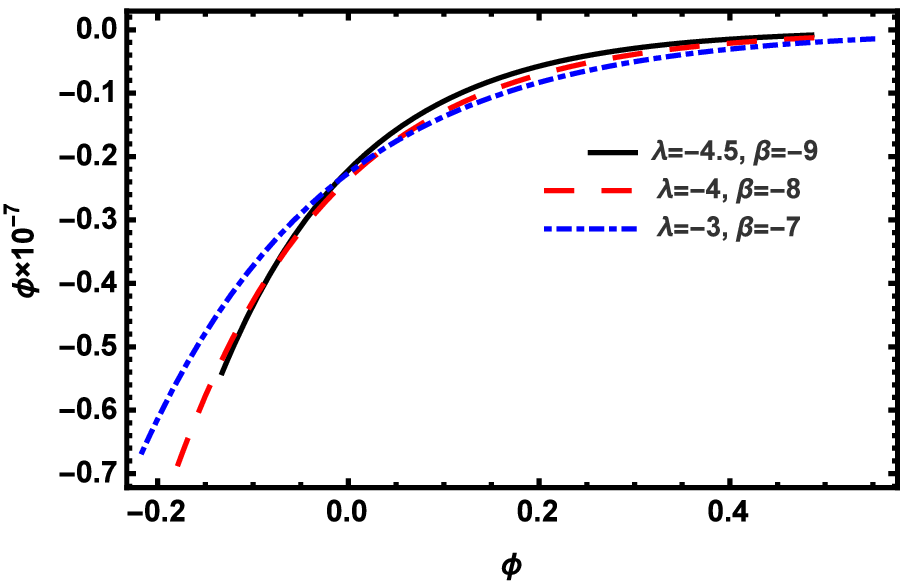}}
  \subfigure[\label{TCepsilon}]{\includegraphics[width=7cm]{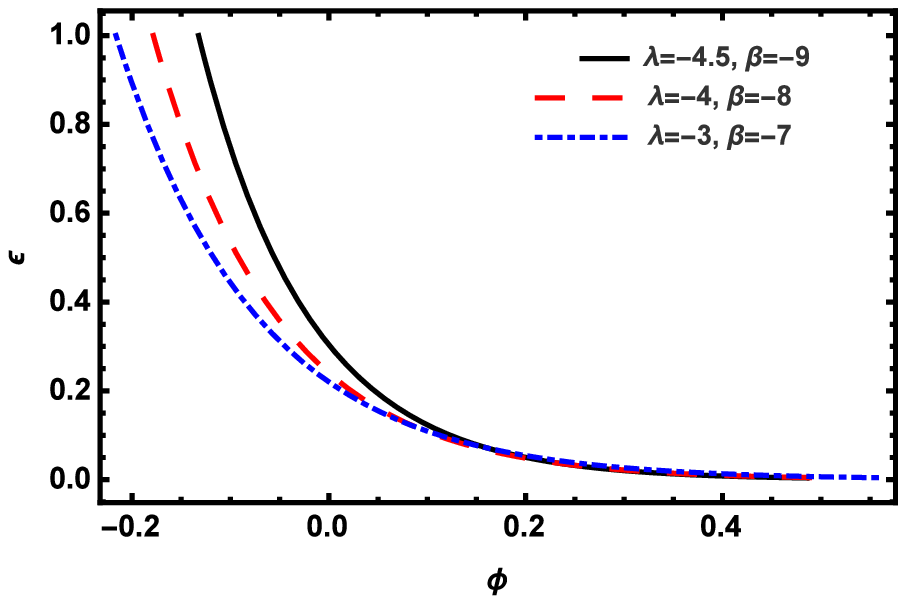}}
  \caption{Behavior of the a) time derivative of the scalar field, b) first slow-roll parameter are plotted versus the scalar field for three different values of $\lambda$ and $\beta$.}\label{TCdphiepsilon}
\end{figure}

\begin{figure}
  \centering
  \includegraphics[width=7cm]{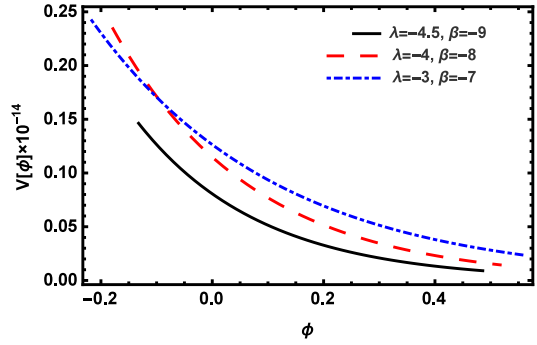}
  \caption{The potential versus the scalar field in inflationary times for three different values of $\lambda$ and $\beta$.}\label{TCpotential}
\end{figure}

More understanding about the results achieved by noting the numerical result which has been prepared in Table.\ref{TCtable}. For each row, a pair of $(\lambda,\beta)$ is taken from the parametric space of Fig.\ref{TClbSC} and for this choice of $\lambda$ and $\beta$ the other parameters are extracted. The results indicate that the model is consistent with observational data, and on the other hand it could successfully satisfies the swampland criteria. However, for the last two rows of the table, the pairs of $(\lambda,\beta)$ is taken from parametric space of Fig.\ref{TClb} in which they are not in parametric space of Fig.\ref{TClbSC}. Although the results about the scalar spectral index and the tensor-to-scalar ratio agree with data, they do not satisfy the first swampland criterion.
\begin{table}
  \centering
  \begin{tabular}{p{1.2cm}p{1.2cm}p{2.5cm}p{2.5cm}p{1.5cm}p{1.5cm}p{1.2cm}p{1cm}}
    \hline
    \ \ $\lambda$ & \ \ $\beta$ & \qquad $V_0$ & \qquad $V^\star$ & \quad $n_s$ & \quad $r$ & $\Delta\Phi$ & $|V'/V|$ \\

    \hline

    $-4.5$ & $-8.0$ & $1.51 \times 10^{-15}$ & $1.51 \times 10^{-16}$ & $0.9626$ & $0.0110$ & $0.712$ & \ \ $4.5$ \\

    \ \ $4.5$ & \ \ $8.0$ & $1.51 \times 10^{-15}$ & $1.51 \times 10^{-16}$ & $0.9626$ & $0.0110$ & $0.712$ & \ \ $4.5$ \\

    $-4.0$ & $-9.0$ & $6.19 \times 10^{-16}$ & $8.72 \times 10^{-17}$ & $0.9608$ & $0.0119$ & $0.600$ & \ \ $4.0$ \\

    \ \ $4.0$ & \ \ $9.0$ & $6.19 \times 10^{-16}$ & $8.72 \times 10^{-17}$ & $0.9608$ & $0.0119$ & $0.600$ & \ \ $4.0$ \\

    $-3.0$ & $-7.0$ & $1.26 \times 10^{-15}$ & $2.35 \times 10^{-16}$ & $0.9606$ & $0.0120$ & $0.776$ & \ \ $3.0$ \\

    \ \ $3.0$ & \ \ $7.0$ & $1.26 \times 10^{-15}$ & $2.35 \times 10^{-16}$ & $0.9606$ & $0.0120$ & $0.776$ & \ \ $3.0$ \\

    \hline

    $-2.0$ & $-5.0$ & $2.27 \times 10^{-15}$ & $8.89 \times 10^{-16}$ & $0.9601$ & $0.0122$ & $1.076$ & \ \ $2.0$ \\

    $-1.0$ & $-1.5$ & $2.07 \times 10^{-13}$ & $1.32 \times 10^{-13}$ & $0.9641$ & $0.0102$ & $3.970$ & \ \ $1.0$ \\
    \hline
  \end{tabular}
  \caption{table}\label{TCtable}
\end{table}
The best choices of the free constants of the case is presented in Fig.\ref{TClbSC}, and by utilizing them one could specify the behavior of the parameters $\dot{\phi}$ and $\epsilon$ and check their consistency during inflation. The parameters are depicted versus the scalar field in Fig.\ref{TCdphiepsilon}. There is the same behavior for $\dot\phi$ and $\epsilon_1$ as we had in the previous section. The parameter $\dot{\phi}$ is negative and the scalar field decreases during inflation. The first slow-roll parameter $\epsilon_1$ is small for the bigger values of the scalar field and it increases and reaches one for the smaller values. The scalar field crosses from the positive values to the negatives values by approaching to the end of inflation. Negative values of the scalar field is not unacceptable but, in general, it is not desirable amongst scientists. \\
The potential of the case is presented in Fig.\ref{TCpotential} for different choices of $\xi$ and $f_0$. In contrast to the previous cases, the potential stands in smaller values at the beginning of inflation and it increases by passing time and approaching to the end of inflation. This is in direct tension with our usual definition and exception of the slow-roll inflation. It is usually stated that during inflation the scalar field slowly rolls down from the top of the potential to the minimum, namely the potential decreases as approaching to the end of inflation. However, here the potential is increasing and it might make it difficult to have the particle production in the reheating phase; in which in the standard model the particle production is produced when the scalar field reaches to the minimum of its potential, its kinetic energy increases, and starts oscillating. This behavior of the potential is strange and not acceptable for the scenario of inflation.

\section{Reheating Era}
\begin{figure}[h]
  \centering
  \subfigure[\label{FCNr}]{\includegraphics[width=7cm]{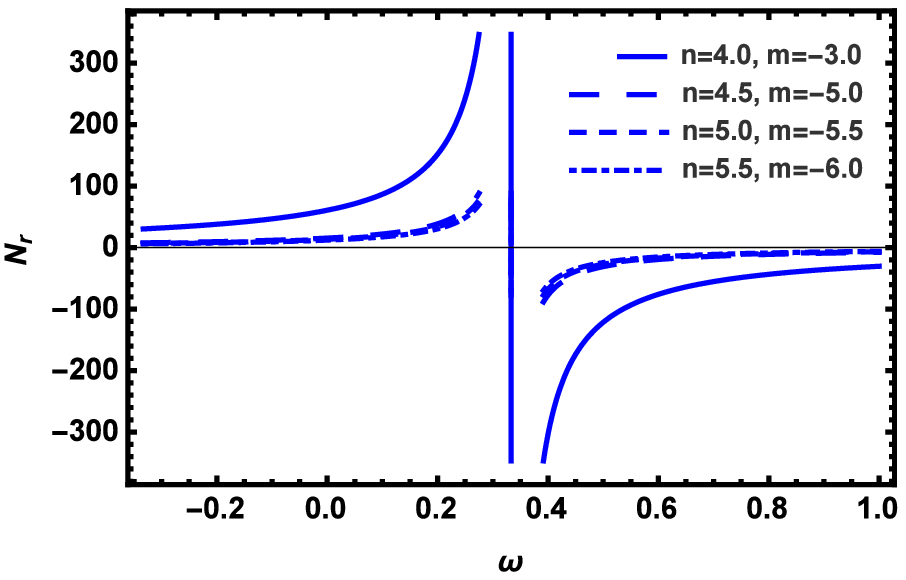}}
  \subfigure[\label{FCTr}]{\includegraphics[width=7cm]{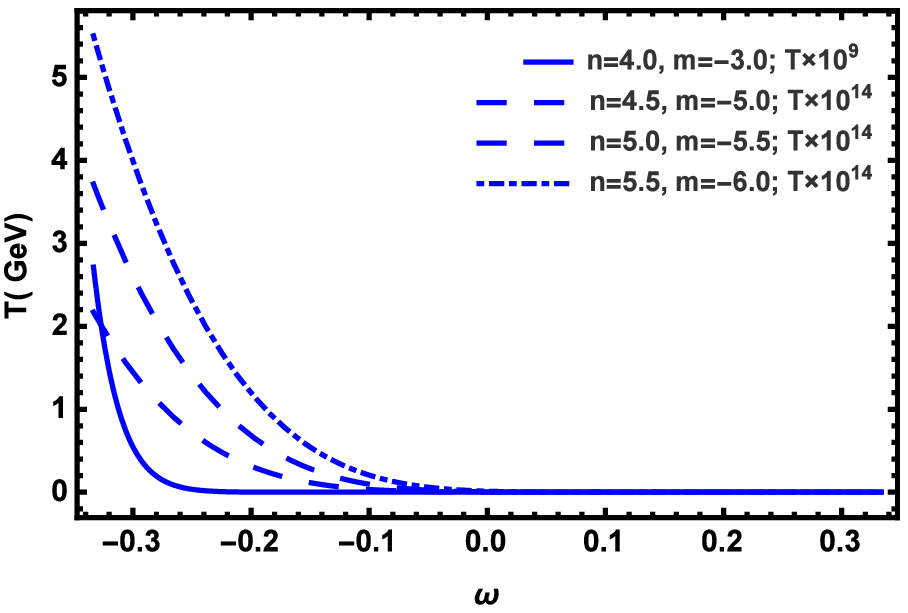}}
  \caption{Behavior of the reheating a) number of e-fold, b) temperature are plotted versus the parameter $\omega$ for three different values of $n$ and $m$.}\label{FCNrTr}
\end{figure}

\begin{figure}[h]
  \centering
  \subfigure[\label{SCNr}]{\includegraphics[width=7cm]{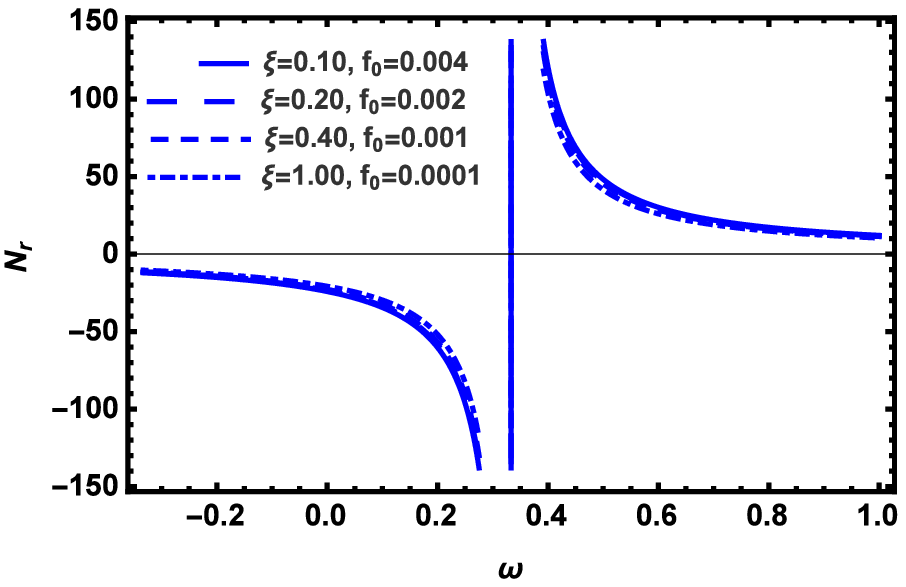}}
  \subfigure[\label{SCTr}]{\includegraphics[width=7cm]{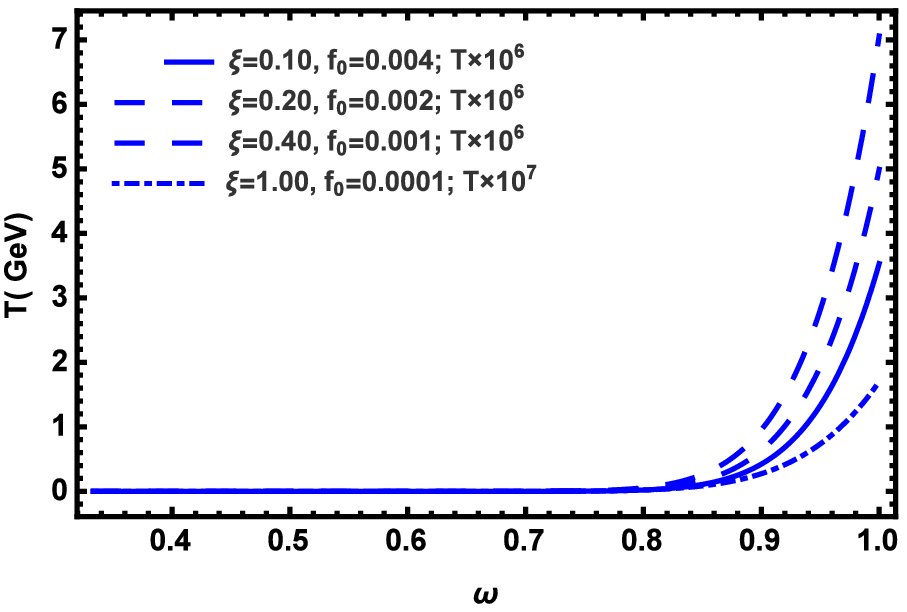}}
  \caption{Behavior of the reheating a) number of e-fold, b) temperature are plotted versus the parameter $\omega$ for three different values of $\xi$ and $f_0$.}\label{SCNrTr}
\end{figure}

During inflation, the universe stands in the accelerated expansion phase and it undergoes an expansion about $N_k$ e-fold from the time of the horizon crossing of the perturbation with wavenumber $k$ to the end of inflation. After the end of inflation, the universe enters to the reheating ear where the stored energy in the scalar field decays to the other particles and it heats up. The universe is no longer in the accelerated expansion phase but it is expanding and experiences about $N_{r}$ e-fold expansion. Due to the decay of energy from the scalar field to the other particles and the interaction between particles, the universe warms up and reaches to a temperature $T_r$. Then, the universe goes to the radiation dominant era and the expansion continues in which the universe expands for another $N_{RD}$ e-fold up to the end of the phase, and then the matter dominant phase begins. This process leads to the following relation which somehow displays the expansion of the universe from the time of the horizon crossing to the present time where we can observe the corresponding perturbation which exits the horizon at the time $t_i$ in inflationary times \cite{Martin:2010kz,Liddle:2003as,Dai:2014jja,Cook:2015vqa,Bhattacharjee:2016ohe,Drewes:2017fmn}
\begin{equation}\label{totalN}
  N = \ln\left( {a_{0} \over a_{\star}} \right) = \ln\left( {a_{0} \over a_{rec}} \right) + \ln\left( {a_{rec} \over a_{r}} \right) + \ln\left( {a_{r} \over a_{end}} \right) + \ln\left( {a_{end} \over a_{\star}} \right)
\end{equation}
where $a_{rec}$ is the scale factor at the recombination time. The subscripts "r" and "end" stands for the time of the end of reheating phase and end of inflation, respectively. In this section, we are interested in the reheating phase in which the number of e-fold at this period of time is described as
\begin{equation}\label{Nr}
  N_r = \ln\left( {a_{r} \over a_{end}} \right).
\end{equation}
Following some routine process (presented in refs.\cite{Liddle:2003as,Dai:2014jja,Cook:2015vqa,Bhattacharjee:2016ohe,Drewes:2017fmn}), one could obtain the number of e-fold of reheating and its temperature based on the constants of the model. Here, we briefly present the procedure and more detail could be found in \cite{Liddle:2003as,Dai:2014jja,Cook:2015vqa,Bhattacharjee:2016ohe,Drewes:2017fmn}. After the inflation, the universe exits from the accelerated expansion phase and the effective equation of state parameter $\omega$ (which is ratio of pressure to energy density) should be larger than $\omega \geq -1/3$ indicating end of inflation. It is assumed that the effective equation of state parameter during inflation is almost constant. Applying this assumption, the energy density in the reheating phase is expressed in terms of the scale factor as follows
\begin{equation}\label{rhor}
  \rho_r = \rho_{end} \; \left( {a_{end} \over a_r} \right)^{3(1+\omega)}
\end{equation}
where $\rho_{end}$ is known as the energy density at the end of inflation and $\rho_r$ is the energy density at the end of reheating era. Substituting this relation in Eq.\eqref{Nr}, leads to
\begin{equation}\label{Nrrho}
  N_r = {1 \over 3(1+\omega)} \; \ln\left( {\rho_{end} \over \rho_r} \right)
\end{equation}
The energy density $\rho_r$ is related to the temperature $T_r$ as follows
\begin{equation}\label{rhoT}
  \rho_r = {\pi^2 g_{re} \over 30} \; T_r^4
\end{equation}
where $g_{re}$ is known as the effective number of relativistic species. After some straightforward manipulation, the reheating temperature and the number of e-fold are obtained as
\begin{eqnarray}
  N_r &=& {-4 \over (1-3\omega)} \; \left[ {1 \over 4} \; \ln\left( {30 \over \pi^2 g_{re}} \right) +
                                           {1 \over 3} \; \ln\left( {11 g_{re} \over 43} \right) +
                                           \ln\left( {k \over a_0 T_0} \right) + \ln\left( {\rho_{end} \over H_k} \right)  + N_k \right] \\
  T_r &=& \left( {43 \over 11 g_{re}} \right)^{1 / 3} \; {a_0 T_0 \over k} \; H_k \; e^{-N_k} \; e^{N_r}
\end{eqnarray}
where $a_0$ is the scale factor at the present time, $H_k$ is the Hubble parameter when the perturbation with wavenumber $k$ exits the horizon, and $N_e$ is measure the e-fold expansion of the universe from the crossing time to the end of inflation. The wavenumber $k$ is the pivot scale and by following Planck is taken as $k/a_0 = 0.005 {\rm MpC}^{-1}$.  \\
Based on our discussion in the previous case, we could determine the free constants of the model using the observational data for number of e-fold $N_k=65$. Using this determined constants, one estimated the reheating number of e-fold and temperature. Figs.\ref{FCNrTr} illustrates the number of e-fold and temperature of the reheating versus the parameter of the equation of state for the first case of the previous section. It is realized from Fig.\ref{FCNr} that for $-1/3 < \omega < 1/3$, the number of e-fold $N_r$ is positive but for $1/3 < \omega < 1$ it is negative and unacceptable for us. Then, in Fig.\ref{FCTr} the temperature behavior is plotted only for $-1/3 < \omega < 1/3$. However, only when the parameter $\omega$ tends to $\omega \rightarrow -1/3$, the $N_r$ could be small and as $\omega$ gets larger $N_r$ gets bigger in which we could have a huge number of e-fold. Our common belief is that the universe undergoes a few number of e-fold during the reheating phase. Therefore, to achieve this belief, the equation os state during the reheating phase should be very close to $-1/3$. The reheating temperature for this value of the effective equation of state is about $10^{14} {\rm GeV}$.
Figs.\ref{SCNrTr} illustrates the number of e-fold and temperature of the reheating versus the parameter of the equation of state for the second case. Unlike the first case, the number of reheating is positive for $1/3<\omega<1$ and it is negative for $-1/3<\omega<1/3$, as shown in Fig.\ref{SCNr}. Since the negative number of e-fold is not acceptable for us, the temperature is plotted for $1/3<\omega<1$. The universe experiences a few number of e-folds as $\omega$ tends to $1$ and it undergoes larger expansion as $\omega$ approaches $1/3$. Since large number of e-fold is not desirable for the reheating phase, it is assumed that the effective equation of state parameter is close to $1$ during the reheating phase. The reheating temperature under this condition is about $10^7 {\rm GeV}$.

\section{Conclusion}
The scenario of inflation in the frame of scalar-tensor theory and its simultaneous consistency with the observational data and swampland criteria was studied. After deriving the basic dynamical and perturbation equations, more investigation and detail were followed by selecting three different cases for the potential and the coupling function. The first step for each case was considering the consistency of the model with observational data. In this regard, the perturbation parameters were estimated at the time of the horizon crossing, which determined that they depend on the free constants of the model. Then, using the $r-n_s$ diagram of Planck and by employing programming we achieved to obtain a parametric space in which for every point in this space the model comes to an agreement with observational data. The data for the amplitude of the scalar perturbations and in some cases the energy scale of inflation were utilized to determine the other constants of the model which in general they also depend on the constants of the parametric space. \\
By comparing the model with data, we found a range for the free constants of the model to verify the model validation. Next, we sought to found whether, for all points in this parametric space, the model could satisfy the swampland criteria as well. The results clarified that only for part of the parametric space the swampland criteria are satisfied, and for the other part at least one of the criteria is violated. In other words, the swampland criteria impose another limitation on the acceptable ranges of the free constants of the model. Combining the observational constraint and the swampland criteria, a parametric space of the free constants of the model was found in which for every point in this space the model simultaneously agrees with observational data and the swampland criteria. \\
Then, the behavior of time derivative of the scalar field, first slow-roll parameter, and the potential were illustrated to considered self-consistency of the model. For all cases, the time derivative of the scalar field was negative, indicating that the scalar field gets smaller by approaching to the end of inflation. The slow-roll parameter $\epsilon_0$ has smaller values for larger field, and it gets larger by reducing inflation. Also, the potential has a smooth behavior in which it is large at the beginning of inflation and decreases by approaching the end of inflation, which states that the scalar field slowly rolls down its potential toward the minimum. The strange situation was happen for the third case, where by approaching to the end of inflation, the potential increase, instead of the usual decreasing behavior.  \\
In the last part of the work, the reheating phase was considered to get some information about the number of e-fold that the universe experiences during the phase and also the reheating temperature. The behavior of the parameters were depicted versus the parameter of equation of state. The result indicates that the parameter $\omega$ should be very close to $\omega \simeq -1/3$ for the first case and $\omega \simeq 1$ for the second case to have desirable values for $N_r$ and reheating temperature, $T_r$.

\bibliography{STTref}

\end{document}